\documentstyle[11pt,paspconf]{article}

\def\lya{Ly$\alpha$~}

\begin{document}

\title{The First Stars and Quasars in the Universe}
\author{Abraham Loeb}
\affil{Astronomy Department, Harvard University, 60 Garden Street,
Cambridge, MA 02138}

\begin{abstract}

The transition between the nearly smooth initial state of the Universe and
its clumpy state today occurred during the epoch when the first stars and
low-luminosity quasars formed. For Cold Dark Matter cosmologies, the
radiation produced by the first baryonic objects is expected to ionize the
Universe at $z\approx 10$--$20$ and consequently suppress by $\sim 10\%$
the amplitude of microwave anisotropies on angular scales $\la 10^\circ$.
Future microwave anisotropy satellites will be able to detect this
signature. The production and mixing of metals by an early population of
stars provides a natural explanation to the metallicity, $\sim 1\%Z_\odot$,
found in the intergalactic medium at redshifts $z\la 5$.  The Next
Generation Space Telescope (NGST) will be able to image directly the
``first light'' from these stars.  With its nJy sensitivity, NGST is
expected to detect $\sim 10^3$ star clusters per square arcminute at $z\ga
10$.  The brightest sources, however, might be early quasars.  The infrared
flux from an Eddington luminosity, $10^6M_\odot$, black hole at $z=10$ is
$\sim 10$ nJy at $1\mu$m, easily detectable with NGST. The time it takes a
black hole with a radiative efficiency of $\sim 10\%$ to double its mass
amounts to more than a tenth of the Hubble time at $z=10$, and so a fair
fraction of all systems which harbor a central black hole at this redshift
would appear active.  The redshift of all sources can be determined from
the Lyman-limit break in their spectrum, which overlaps with the NGST
wavelength regime, 1--$3.5\mu$m, for $10<z<35$.  Absorption spectra of the
first generation of star clusters or quasars would reveal the reionization
history of the Universe.  The intergalactic medium might show a significant
opacity to infrared sources at $z\ga 10$ due to dust produced by the first
supernovae.

\end{abstract}

\section{Introduction}

The cosmic microwave background (CMB) anisotropies detected by the COBE
satellite (Bennet et al. 1996) confirmed the notion that the present
structure in the Universe originated from small density fluctuations at
early times.  The gravitational collapse of overdense regions could explain
the present-day abundance of bound objects, such as galaxies or X-ray
clusters, under the appropriate extrapolation of the detected large-scale
anisotropies to smaller scales (e.g., Baugh et al. 1997; Pen 1996). Recent
deep observations with the Hubble Space Telescope (Steidel et al. 1996;
Madau et al. 1996) and ground-based telescopes, such as Keck (Lowenthal et
al. 1996), have constrained considerably the evolution of galaxies and
their stellar content at $z\la 5$.  However, in the bottom-up hierarchy of
the popular Cold Dark Matter (CDM) cosmologies, galaxies were assembled out
of building blocks of smaller mass. The elementary building blocks, i.e.
the first gaseous objects to form, had a total mass of order the Jeans mass
($\sim 10^{6}M_\odot$), below which gas pressure opposed gravity and
prevented collapse (Haiman \& Loeb 1997; Ostriker \& Gnedin 1997). In the
standard CDM cosmology, these {\it basic} building blocks formed at
$z\sim 10$--$30$ (see Fig. 1).

\begin{figure}
\includegraphics{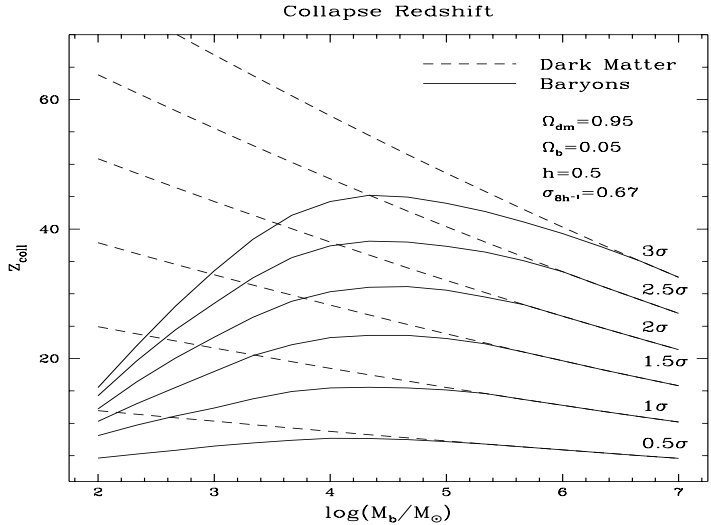}
\vspace{2.7in}
\caption{Collapse redshift for cold dark matter (dashed lines) 
and baryons (solid lines) in spheres of various masses and initial
overdensities. The overdensities are in units of the rms amplitude of
fluctuations $\sigma(M)$ for a standard CDM power-spectrum with
$\sigma_{8h^{-1}}=0.67$. The collapse of the baryons is delayed relative to
the dark matter due to gas pressure.  The curves were obtained by following
the motion of the baryonic and dark matter shells with a spherically
symmetric, Lagrangian hydrodynamics code (Haiman \& Loeb 1997).  }
\label{fig-1}
\end{figure}

The {\it first light} from stars and quasars ended the ``dark ages'' of the
universe and initiated a ``renaissance of enlightenment'' in the otherwise
fading glow of the big bang. It is easy to see why the mere conversion of
trace amounts of gas into stars or black holes at this early epoch could
have had a dramatic effect on the ionization state and temperature of the
rest of the gas in the Universe.  Nuclear burning releases $5\times 10^6$
eV per hydrogen atom, and thin-disk accretion onto a Schwarzschild black
hole releases ten times more energy; however, the ionization of hydrogen
requires only 13.6 eV.  It is therefore sufficient to convert a minimum
fraction of $\sim 10^{-5}$ of the baryonic mass into stars or black holes
in order to ionize the rest of the Universe. (The actual required fraction
is higher because some of the emitted photons are below 13.6 eV and because
each hydrogen atom recombines multiple times at high redshifts).  The free
electrons produced in this process erase the microwave background
anisotropies on angular scales below the size of the horizon at the
reionization epoch ($\sim 10^\circ$), by an amplitude comparable to their
optical depth to Thomson scattering.

A variety of CDM models that  are all consistent with both the COBE
anisotropies ($z\approx 10^3$) and the abundance of objects today ($z=0$)
differ appreciably in their initial amplitude of density fluctuations on
small scales.  The reionization history of the Universe is determined by
the collapse redshift of the smallest objects ($\sim 10^{6-9}M_\odot$) and
is therefore ideally suited to discriminate among these models.

We have not yet observed the first transition that the Universe made
between its nearly smooth initial state and its clumpy present state. If I
had been asked to envision an instrument that would directly probe this
transition epoch, I would have probably sketched something like the Next
Generation Space Telescope (NGST). There are three reasons that make this
choice natural: (i) At $z\sim 10-50$, the rest frame optical--UV emission
of star clusters or quasars is redshifted to infrared wavelengths in the
observer frame; (ii) The first sources are faint because of their high
redshift and low mass. Their detection, therefore, requires a sensitive
infrared telescope with a large collecting area, in space. The last
requirement is necessary to avoid atmospheric background/opacity and
to reduce the zodiacal background through an extended solar orbit; and
(iii) High-redshift objects are likely to be denser and more compact than
their low-redshift counterparts. This simply follows from the increase in
the mean density of the Universe at early times. In the bottom-up hierarchy
of structure formation, this tendency is also enhanced by the decrease of
the nonlinear mass scale with redshift. The physical size of an object of
mass $M$ should scale roughly as $\propto M^{1/3}/(1+z_f)$, where $z_f$ is
its formation redshift. High angular resolution is therefore required for
imaging star clusters at $z\ga 10$.

The detection of the transition epoch could provide invaluable information
about the ionization and star formation history of the Universe. The study
of stars at high-redshift is also linked to the ``archeology'' of old stars
in the Milky Way halo.  The metallicity-age trends which are identified in
the Milky-Way galaxy (McWilliam et al. 1996) should be related to the
various epochs of star formation in the early Universe.

So much for theoretical arguments.  But is there direct evidence that
structure started to form in the Universe long before galaxies were
assembled? The answer is, yes.  By now, a number of observational clues
constitute a ``smoking gun'' for an early population of stars that had
formed at $z\ga 5$:

\begin{itemize}

\item Recent spectroscopic observations of the \lya forest at $z\ga 3$ show
evidence for a metallicity $\sim 1\%Z_\odot$ in absorbers with HI column
densities as low as $10^{15}~{\rm cm^{-2}}$ (Cowie~1996;
Songaila~\&~Cowie~1996; Tytler~et~al.~1995).  Numerical simulations
identify such absorbers with mildly overdense regions in the intergalactic
medium, out of which nonlinear objects such as galaxies condense (Hellstein
et al.  1997, and references therein).  Indeed, damped \lya absorbers,
which are thought to be the progenitors of present--day galaxies, show
similar metallicities during the early phase of their formation, at $3.5\la
z\la 4.5$ (Lu, Sargent, \& Barlow 1996).  The universality of this mean
metal abundance in absorbers, spanning a range of six orders of magnitude
in HI column density, indicates that an early phase of metal enrichment
by stars occurred throughout the Universe at $z\ga 5$.

\item The lack of complete absorption shortward of the
\lya~resonance (the so-called ``Gunn-Peterson effect''; Gunn \& Peterson
1965) in quasar spectra at $z\sim 5$ implies that the intergalactic gas was
highly photoionized before that time. On the other hand, the bright quasar
population has been observed to decline at $z\ga 3$, and so stellar sources
might be necessary to keep the intergalactic gas in its inferred
high-ionization state (Shapiro \& Giroux 1987; Miralda-Escud\'e \& Ostriker
1990).

\item Microlensing surveys argue that $\sim 50^{+30}_{-20}\%$ 
(instead of the previously known range of 0--100\%!) of the mass of the
halo of the Milky Way galaxy might be in the form of Massive Compact Halo
Objects (MACHOs) with a mass $\sim 0.5^{+0.3}_{-0.2} M_\odot$ (Alcock et
al. 1996). It remains to be seen whether future microlensing studies
confirm these early reports. A stellar population that forms before
galaxies are assembled, would behave as collisionless matter and populate
the halos of galaxies like Cold Dark Matter (CDM) particles do.

\end{itemize}

In this review I will describe the properties of the first stars and quasars
and their effect on the ionization history of the Universe. Much of the
quantitative details of this sketch can be found in a series of papers that
I wrote in collaboration with my student, Zoltan Haiman.  The underlying
theme of these papers was to calibrate the net fraction of gas that is
converted into stars based on the measured C/H ratio in the
\lya forest. This calibration fixes the reionization history 
for a given stellar Initial Mass Function (IMF) and a particular CDM
cosmology.

\section{Properties of the First Star Clusters}

The various stages in the reionization history of the Universe are
illustrated in Figure 2. This sequence follows the collapse redshift
history of baryonic objects shown in Figure 1, which was calculated with a
spherically-symmetric code for gas and dark matter (Haiman, Thoul, \& Loeb
1996). For objects with baryonic masses $\ga 3\times 10^4M_\odot$, gravity
dominates and results in the characteristic bottom-up hierarchy of CDM
cosmologies.  At lower masses, gas pressure delays the collapse and results
in a top-down hierarchy, reminiscent of hot dark matter cosmologies. The
first objects to collapse are located at the ``knee'' that separates these
regimes.  Such objects reach virial temperatures of several hundred degrees
and could fragment into stars only through cooling by molecular hydrogen
[see Haiman et al. (1996) or Tegmark et al. (1997), for details regarding
the chemistry network leading to the formation of ${\rm H_2}$ in a
primordial gas].

\begin{figure}
\includegraphics{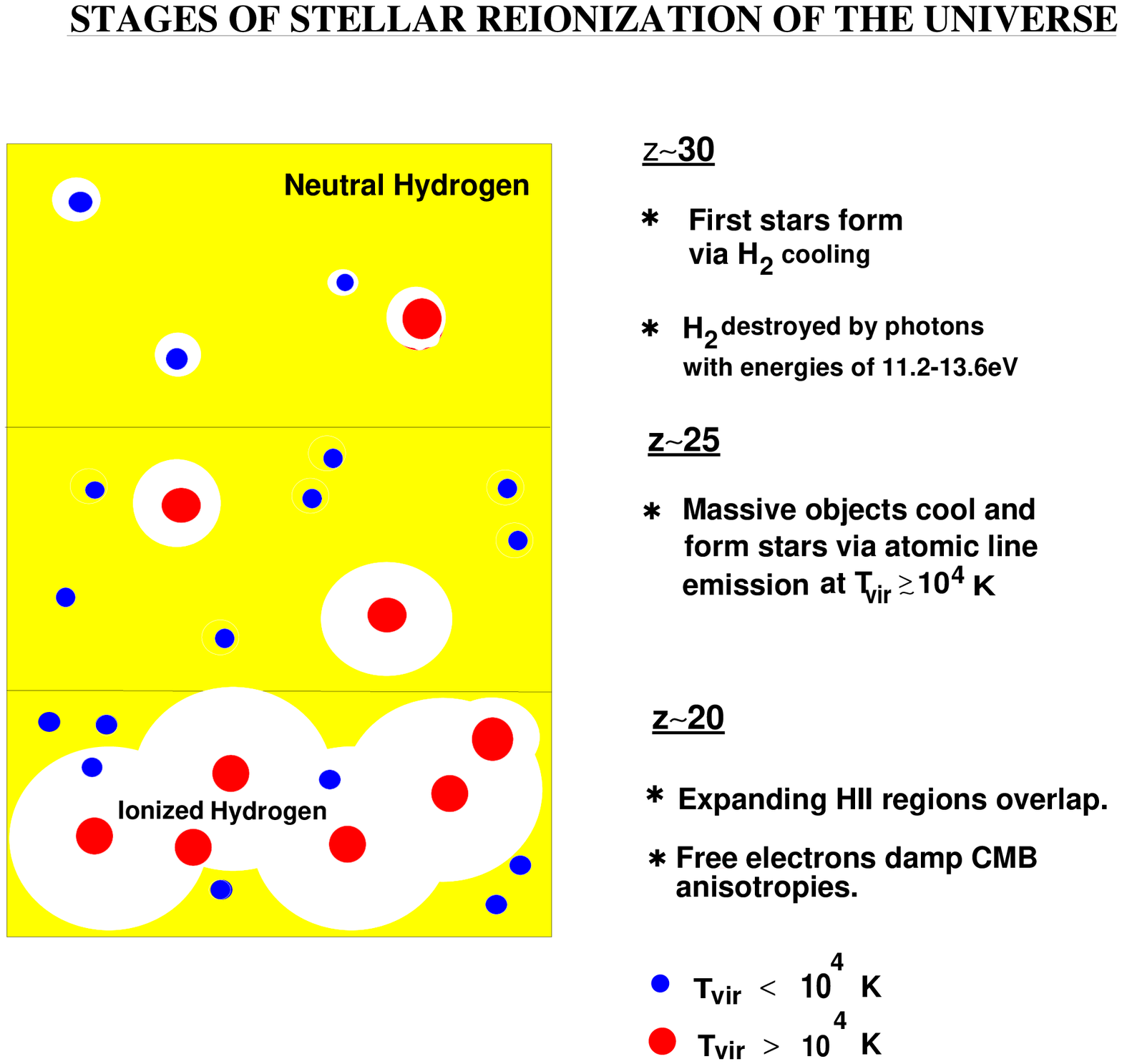}
\vspace{3.in}
\caption{} 
\label{fig-2}
\end{figure}

However, ${\rm H_2}$ is fragile and could easily be photo-dissociated by
photons with energies of $11.2$--$13.6$eV, to which the Universe is
transparent even before it gets ionized. Haiman, Rees, \& Loeb (1996)
showed that a UV flux of $\la 1~{\rm erg~cm^{-2}~s^{-1}~Hz^{-1}~sr^{-1}}$
is capable of dissociating ${\rm H_2}$ throughout the collapsed
environments in the Universe.  This flux is lower by $\ga 2$ orders of
magnitude than the minimum flux necessary to ionize the Universe, which
amounts to one UV photon per baryon.  The inevitable conclusion is that
soon after the first stars form, the formation of more stars due to ${\rm
H_2}$ cooling is suppressed.  Further fragmentation is possible only
through atomic line cooling, which is effective in objects with high virial
temperatures, $T_{\rm vir}\ga 10^4$K.  Such objects correspond to a total
mass $\ga 10^8 M_\odot [(1+z)/10]^{-3/2}$. Figure 2 illustrates this
sequence of events by describing two classes of objects: those with $T_{\rm
vir}< 10^{4}$K (small dots) and those with $T_{\rm vir}> 10^4$K (large
dots).  In the first stage (top panel), some low-mass objects collapse,
form stars, and create ionized hydrogen (HII) bubbles around them.  Once
the UV background between 11.2--13.6eV reaches some critical level, ${\rm
H_2}$ is photo-dissociated throughout the Universe and the formation of new
stars is delayed until objects with $T_{\rm vir}\ga 10^4$K collapse.  Each
massive source creates an HII region which expands into the intergalactic
medium.  Initially the volume of the Universe is dominated by neutral
hydrogen (HI).  But as new sources appear exponentially fast (due to the
Gaussian nature of the perturbations), numerous HII bubbles add up,
overlap, and transform all the remaining HI into HII over a short period of
time. This process resembles a first-order phase transition.  Since the
characteristic separation between sources is eventually very much smaller
than the Hubble distance, the transition completes over a period of time
which is much shorter than the Hubble time, and can be regarded as sudden.

Because the potential wells of the first clusters are relatively shallow
($\sim$$10~{\rm km~s^{-1}}$), supernova--driven winds are likely to have
expelled the metal--rich gas out of these systems and mixed it with the
intergalactic medium.  Incomplete mixing could have led to the observed
order-of-magnitude scatter in the C/H ratio along lines-of-sight to
different quasars (Rauch, Haehnelt, \& Steinmetz 1996;
Hellsten~et~al.~1997). It is an interesting coincidence that the supernova
energy output associated with a metal enrichment of $\sim 1\%Z_\odot$
corresponds to $\sim 10$ eV per hydrogen atom, which is just above the
binding energy of these early star clusters. Supernova feedback in these
objects could have therefore dictated the average metallicity observed in
the \lya forest. Direct observations of these supernovae might be feasible
in the future (Miralda-Escud\'e \& Rees 1997).

\begin{figure}
\includegraphics{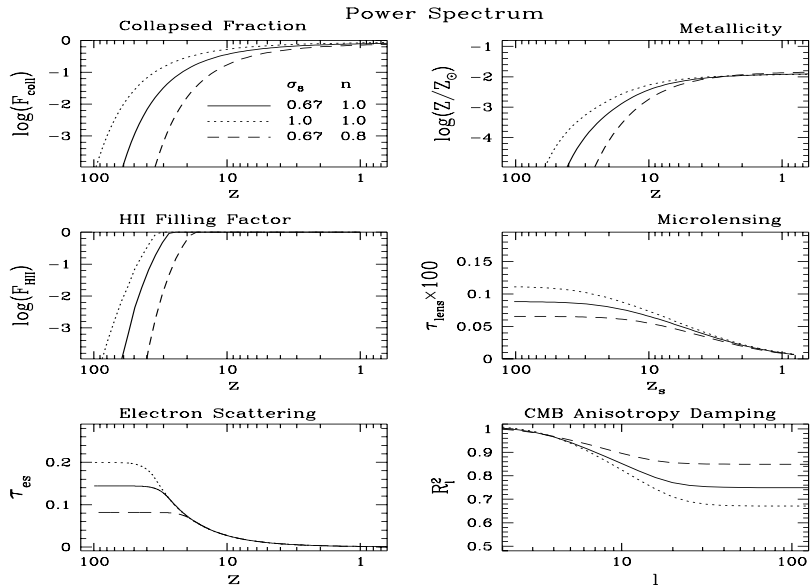}
\vspace{2.95in}
\caption{The different panels show: (i) the collapsed
fraction of baryons; (ii) the average metallicity of the IGM; (iii) the
volume filling factor of ionized hydrogen; (iv) the optical depth to
cosmological microlensing; (v) the optical depth to electron scattering;
and (vi) the net damping factor for the power-spectrum decomposition of
microwave anisotropies as a function of the spherical harmonic index $l$.
The different curves show the sensitivity of the results to changes in the
normalization $\sigma_{8h^{-1}}$ or the primordial index $n$ of the CDM
power-spectrum.}
\label{fig-3}
\end{figure}

The measured C/H ratio can be used to calibrate the net fraction of gas
that is converted into stars by a redshift $z\approx3$. For a Scalo (1986)
IMF of stars, this corresponds to converting $\sim 3\%$ of all the gas into
stars\footnote{There is a remaining uncertainty of whether $3\%$ of the gas
in each object formed stars or whether a small fraction of all objects
formed stars with a higher efficiency.  We assume that supernova feedback
limits star formation to a single starburst and sets the efficiency of star
formation to $3\%$ in all objects. In the alternative scenario, NGST should
see fewer but brighter sources.}. Haiman \& Loeb (1997) have used this
fraction and an extension of the Press--Schechter formalism to calculate
the early star formation history in a variety of CDM cosmologies.  We
combined the collapse redshift information in Figure 1 with the
Press-Schechter theory and assumed that each object converts a fixed
fraction of its gas into stars in a single starburst.  We then used
the spectral atlases of Kurucz (1993) and the stellar evolutionary tracks of
Schaller et al. (1992) to find the emission spectrum of the stars as a
function of redshift. The ionizing radiation from each star cluster was
absorbed in part by the gas in it (assumed to be distributed as a singular
isothermal sphere at the virial temperature) and then propagated into a
smooth intergalactic medium (IGM), using the radiative transfer equations in
an expanding universe.

For a wide range of models and a Scalo IMF, we have found that the Universe
is reionized by a redshift $z=10$--20, and that the resulting optical depth
to electron scattering is $\sim 10\%$ (cf. Fig. 3).  This result changes
qualitatively only if the IMF is tilted by a power-law index $\ga 1$
relative to the Scalo shape (a bias towards high mass stars is likely due
to the lack of cooling by metals and the associated increase in the Jeans
mass).  The sensitivity to model parameters other than the IMF is only
logarithmic because of the exponential dependence of the collapsed baryonic
fraction on redshift for Gaussian fluctuations.  However, the formation
time of CDM halos has a very weak dependence on mass scale at low masses,
and it is not obvious whether the Press-Schechter approach is accurate in
this limit.  Three-dimensional simulations (Gnedin \& Ostriker 1997) find
reionization somewhat later, at $z\approx7$, based on a significantly
different star formation history than the our semi-analytic approach
predicts.  Future microwave anisotropy satellites, such as the Microwave
Anisotropy Probe (MAP) or the Planck Surveyor, could detect a reionization
optical depth as low as 2\% or 0.5\% respectively after processing their
polarization data (Zaldarriaga, Spergel, \& Seljak 1997), and will
therefore test these predictions.

\begin{figure}
\includegraphics{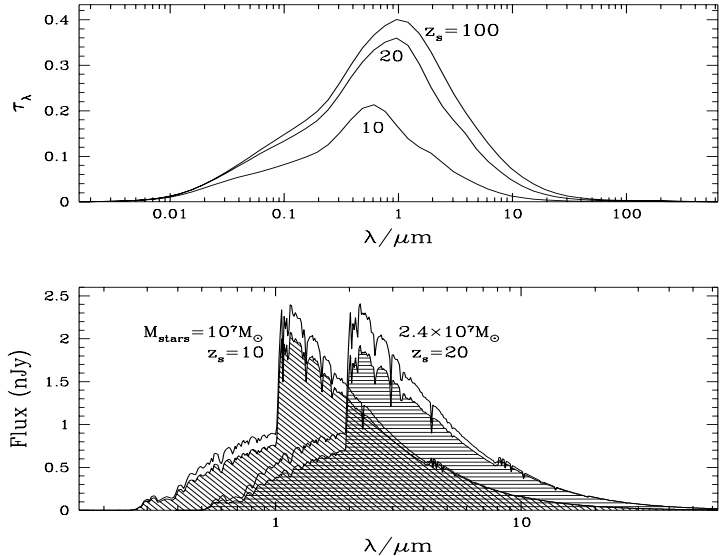}
\vspace{2.35in}
\caption{{\bf Top panel}: optical depth
to absorption and scattering by intergalactic dust at $z\geq3$, as a
function of observed wavelength. The vertical scale is proportional to the
mass of dust produced per supernova in units of $0.3M_\odot$. The source is
located at redshifts $z_{\rm s}$=$10$, $20$, or $100$.  {\bf Bottom panel}:
the expected flux with (lower curve) and without (upper curve) intervening
dust, from star--clusters at $z_{\rm s}$=10 and $20$, containing $10^7~{\rm
M_{\odot}}$ and $2.4\times10^7~{\rm M_{\odot}}$ in stars.  
The stars are distributed on the main sequence according to a Scalo IMF.  
The break in the spectra at the redshifted Lyman limit wavelength includes
only the absorption by neutral hydrogen in the stellar atmospheres.}
\label{fig-4}
\end{figure}

Deep imaging with future infrared telescopes, such as the Space Infrared
Telescope Facility (SIRTF) or the Next Generation Space Telescope (NGST),
would be able to image pre--galactic star clusters. For standard CDM with
$\sigma_{8h^{-1}}=0.67$ and a Scalo IMF, Haiman \& Loeb (1997) predict that
NGST should find $\ga 10^3$ star clusters per square arcminute at $z>10$,
given its planned detection threshold of 1 nJy in the wavelength range of
1--3.5~$\mu$m (Mather \& Stockman 1996). The characteristic separation
between such sources is a few arcseconds, much larger than their typical
size of $\la 0.1^{\prime\prime}$.  

The early epoch of star formation and metal enrichment is inevitably
accompanied by the formation of dust in Type II supernova shells. This dust
has two important observational signatures. First, the absorption of
starlight energy and its re--emission at long wavelengths distorts the
spectrum of the cosmic microwave background (CMB) radiation (Wright et al.
1994, and references therein).  Second, the opacity of the intergalactic
medium to infrared sources at redshifts $z\ga 10$ could be significant.
For these redshifts, infrared in the observer frame corresponds to UV in
the source frame--a spectral regime in which dust absorption peaks.  Dust
obscuration must therefore be considered when predicting the performance of
future infrared telescopes such as NGST.

Loeb \& Haiman (1997a) find that for a Scalo IMF and a uniform mix of metals
and dust with the intergalactic gas, the dust distorts the microwave
background spectrum by a $y$--parameter in the range $(0.02$--$2)\times
10^{-5} (M_{\rm SN}/0.3{\rm M_\odot})$, where $M_{\rm SN}$ is the average
mass of dust produced per supernova.  Note that this range
is not far below the current COBE limit of $1.5\times 10^{-5}$
(Fixsen et al. 1996). The opacity of intergalactic dust to
infrared sources at redshifts $z\ga 10$ is significant, $\tau_{\rm dust}=
(0.1$--$1)\times (M_{\rm SN}/0.3{\rm M_\odot})$, and could be detected with
NGST.  Although dust suppresses the Ly$\alpha$ emission from early sources,
the redshifts of star clusters at $z=10$--35 can be easily inferred from
the Lyman-limit break in their infrared spectrum between 1--3.5$\mu$m (see
Fig. 4).

\section{When Did the MACHOs Form?}

As mentioned before, recent microlensing searches (Alcock et al. 1996)
suggest that $\sim 50^{+30}_{-20}\%$ of the mass of the Milky-Way halo
might be in the form of Massive Compact Halo Objects (MACHOs) with a mass
$\sim 0.5^{+0.3}_{-0.2} M_\odot$.  A substantial MACHO population of dim
stars would naturally be linked to the early universe.  Pre-galactic stars
which originate in abundant low-mass systems are expected to behave as
collisionless particles and populate the halos of galaxies, just as
elementary CDM particles do.  Only stars which are born in massive rare
(high$-\sigma$) peaks are more likely to end up concentrated in galactic
bulges due to the strong dynamical friction which acts on their parent
systems.  In general, dim stars could constitute the dark matter in galaxy
clusters only if $\Omega_b\sim 0.2$ and the efficiency of converting gas
into these stars is of order 80\%.

Recent observations of the star formation history at $z\la 4$ might account
for most of the luminous stellar population in galaxies today (Madau 1996).
However, it is important to remember that this population amounts only to a
density parameter $\Omega_\star\sim 6\times 10^{-3}$ (Woods \& Loeb 1997).
Therefore, if stellar MACHOs make the dark matter with $\Omega\sim 0.2$,
then we have not observed yet $\sim 97\%$ of the star formation history in
the Universe.

NGST could easily detect stellar MACHOs during their earlier history when
they were still bright due to hydrogen burning. For example, Chabrier et
al. (1996) and Adams \& Laughlin (1996) have argued that an old population
of white dwarfs could account for the microlensing data and still be
consistent with metallicity, star counts, and infrared background
constraints, as long as their IMF was highly peaked around $\sim 2M_\odot$.
But a $2M_\odot$ star has a lifetime $\sim 1$ Gyr during which it is more
than an order of magnitude brighter than the Sun. Even if these stars
formed at arbitrarily early times, the age of the Universe is only $\sim
1$Gyr at $z=4$.  Therefore, the halos of galaxies at $z\approx 4$ should be
{\it glowing} in the infrared if dim white dwarfs constitute the dark
matter in the present universe.  Existing Hubble Deep Field observations
(Guzm\'an et al.  1997) already place constraints on the extent of luminous
halos around galaxies (Loeb \& Pinsonneault 1997).  However, optical
observations probe the rest-frame UV and do not see the bulk of the stellar
population.  Infrared observations with NGST would improve dramatically our
ability to set constraints on unusual stellar populations in the halos of
galaxies.

\section{The Next Frontier: Star Formation at High Redshifts}

Most of the current theoretical modeling of star formation was developed to
explain the complex properties of star forming regions in the interstellar
medium (ISM) of our Galaxy. The reason for this bias is obvious: these are
the environments for which data exists!  The situation resembles a search
for a needle in a haystack, just because the haystack happens to be
nearby.  An alternative strategy is to invest (NASA) money in a long-range
mission and search for the needle in simpler environments.

I would like to argue that the formation process of the first generation of
stars in the early Universe should be easier to model than that of stars in
the local ISM.  This follows from the fact that the initial conditions in
cosmology are simpler and better-defined: (i) The Universe started from a
smooth initial state and a specified spectrum of fluctuations on top of it;
many details of this spectrum will be determined by
forthcoming microwave anisotropy experiments; (ii) the composition of the
gas is well defined by standard Big-Bang nucleosynthesis (Schramm
\& Turner 1996), and the chemistry and cooling rates are 
simplified by the lack of heavy elements, dust, or cosmic rays; and
(iii) Magnetic fields were probably negligible before stars formed;
most cosmological scenarios result in small field amplitudes (Sigl et al.
1997, and references therein).

Due solely to lack of data, theoretical studies of the formation
of the first stars are sparse in the literature.  Direct imaging and
spectroscopy of the first stars by NGST could stimulate theoretical work on
the simplest environment for star formation, namely the early Universe.
The era of classical cosmology will end if the MAP and Planck satellites
determine the basic cosmological parameters to a reasonable accuracy.
Since much of observational cosmology relies on starlight, cosmologists
might then find it appropriate to shift their attention to the fundamental
problem of star formation; and there is no better place to start dealing
with it than the early universe.

\section{The First Quasars}

Extrapolation of the quasar luminosity function (Pei 1995) to redshifts
$z\ga 5$ predicts a noticeable tail of faint sources that could be detected
with NGST out to the reionization epoch (Loeb \& Haiman 1997b).  If these
sources are brighter than the first star clusters, it would be easier to
measure their absorption spectra and use them to determine the reionization
redshift beyond which the Gunn-Peterson effect appears.  Bearing in mind
that quasars can account for a considerable fraction of the UV background
at lower redshifts (Meiksin \& Madau 1993), it is quite possible that early
{\it low-luminosity} quasars were in fact responsible for the reionization
of the Universe.  Quasars could be more effective than stars in ionizing
the Universe because: (i) their emission spectrum is harder; (ii) the
radiative efficiency of accretion flows could be more than an order of
magnitude higher than that of a star; (iii) quasars are brighter, and for a
given recombination rate in their host system, the escape fraction of their
ionizing photons is higher than for stars.

First, let us address the black hole formation process.  To form a black
hole inside a given dark matter halo, the baryons must cool.  For most
objects, this is possible only due to atomic line cooling (see \S 2), at
virial temperatures, $T_{\rm vir}\ga 10^4$K, and baryonic masses $\ga
10^7M_\odot [(1+z)/10]^{3/2}$. After losing their thermal pressure, the
cold baryons collapse to a thin disk on a dynamical time (Loeb \& Rasio
1994).  The basic question is then: which fraction of the cold baryons are
able to sink to the very center of the potential well and form a massive
black hole?  The main barrier to this process is angular momentum.  The
centrifugal force opposes radial infall and keeps the gas in typical disks
at a distance which is 6-8 orders of magnitude larger than the
Schwarzschild radius of the system.  However, Eisenstein \& Loeb (1995)
have demonstrated that a small fraction of all objects have a sufficiently
low angular momentum so that the gas in them inevitably forms a compact
semi-relativistic disk that evolves to a black hole on a short viscous
time. These low-spin systems are born in special cosmological environments
that exert unusually small tidal toques on them during their cosmological
collapse.  As long as the initial cooling time of the gas is short and its
star formation efficiency is low, the gas would form this compact disk on a
free-fall time. In typical systems the baryons dominate gravity inside the
scale length of the disk.  Therefore, if the baryons in a low-spin system
acquire a specific angular momentum $j$ which is only a tenth of the
typical value, then the size of the resulting disk ($r\propto j^2$) will be
smaller by a factor of $\sim 10^2$ than average, and the rotation velocity
of the disk will be a factor of $\sim 10$ larger than the velocity
dispersion of the dark matter halo.  For $T_{\rm vir}\sim 10^4$K, the dark
matter halo has a velocity dispersion of $\sim 10~{\rm km~s^{-1}}$, and the
low-spin disk would have a characteristic rotation velocity of $\sim
100~{\rm km~s^{-1}}$ (sufficient to retain the gas against supernova-driven
winds), a size $\la$ pc, and a viscous evolution time shorter than the
Hubble time.  If a low-spin object of this type is embedded inside an
overdense region that will eventually become a galactic bulge, the black
hole progenitor would eventually sink to the center of the bulge by
dynamical friction and seed quasar activity.  Based on the phase-space
volume accesible to low-spin systems ($\propto j^3$), one would expect a
fraction $\sim 10^{-3}$ of all the mass in the Universe to be included in
such systems (Eisenstein \& Loeb 1995).  However, this is a conservative
estimate.  Additional angular momentum loss due to dynamical friction of
gaseous clumps in dark matter halos (Navarro, Frenk,
\& White 1995) or bar instabilities in self-gravitating disks (Shlosman, 
Begelman, \& Frank 1990) could only make the black hole formation process
more prominent.  The popular paradigm that all galaxies harbor black holes
at their center (Haehnelt \& Rees 1993) simply {\it postulates} that in all
massive systems, a small fraction of the gas ends up as a black hole, but
does not explain quantitatively why this fraction obtains its particular
small value.

Observations of galactic nuclei in the local Universe imply black hole
masses which are typically a fraction $\sim 2\times 10^{-3}$ of the total
baryonic mass of their host (see summary of dynamical estimates in Fig. 5
of Kormendy et al. 1997; and the maximum ratio between a quasar luminosity
and its host mass found by McLeod 1997). This ratio is somewhat higher in
low-mass galaxies, such as the compact ellipticals M32 and NGC 4486B.  In
particular, van der Marel et al.  (1997) infer a black hole mass of $\sim
3.4\times 10^6M_\odot$ in M32, which is a fraction $\sim 8\times 10^{-3}$
of the stellar mass of the galaxy, $\sim 4\times 10^8M_\odot$, for the
central mass-to-light ratio of $\gamma_V=2$; while Kormendy et al.  (1997)
infer a black hole mass of $6\times 10^8M_\odot$ in NGC 4486B which is a
fraction $\sim 9\%$ of the stellar mass.

Figure 1 implies that $2\sigma$ peaks with a baryonic mass $\sim
10^{9}M_\odot$ and a velocity dispersion $\sim 100~{\rm km~s^{-1}}$,
similar to M32 or NGC 4486B, collapse at $z\sim 10$ in standard CDM. With a
$3\%$ star formation efficiency, such objects are $\sim 10$nJy stellar
sources for NGST.  We assume that each such object forms a black hole with
an efficiency\footnote{On a given galaxy mass scale, the efficiency of black
hole formation could only increase with increasing formation redshift
$z_f$, since the central density of the host galaxy grows as $\propto
(1+z_f)^3$ while its spin parameter has a negligible dependence on
redshift.} of $10^{-3}$, i.e. the black hole mass is $M_{\rm bh}\sim
10^6M_\odot$. If the black hole shines at a fraction ${\tilde L}$ of the
Eddington limit with a characteristic quasar spectrum, then its infrared
flux at $z\sim 10$ would be $\sim 10 (M_{\rm bh}/10^6M_\odot) {\tilde L}$
nJy. Such a flux is comparable to the emission from the host star cluster
and would easily be detected with NGST.\footnote{Note that although quasars
might be brighter than star clusters and supernovae, they are not
necessarily the brightest sources in the early Universe. The optical
afterglow of $\gamma$-ray bursts could reach much higher luminosities in
the first month$\times(1+z)$ of the burst (Groot et al.  1997; Sahu et al.
1997); however, in the local Universe such events are rarer by $\sim 4$
orders of magnitude than supernovae.}

Among local galaxies, a fraction $\sim 1$--$10\%$ of all galactic nuclei
appear highly active (Huchra \& Burg 1992; Maiolino \& Rieke 1995; Ho,
Fillipenko, \& Sargent 1997; K\"ohler et al.  1997). This fraction can be
explained in terms of the ratio between the Eddington time and the current
Hubble time.
A black hole which shines at a luminosity ${\tilde L}$ in Eddington units
would double its mass on a timescale,
\begin{equation}
t_{\rm Edd}= {M_{\rm bh}\over {\dot M_{\rm bh}}}=
4\times 10^7~{\rm yr}~{\epsilon_{0.1}\over {\tilde L}} ,
\end{equation}
where $\epsilon_{0.1}$ is the radiative efficiency of the black hole in
units of $10\%$. Since the black hole mass, $M_{\rm bh}\propto
\exp\{t/t_{\rm Edd}\}$, grows exponentially with the Eddington $e$-folding
time, the characteristic lifetime of the quasar activity is several times
$t_{\rm Edd}$, namely $\sim 1\% \epsilon_{0.1}$ of the current Hubble time.
But notice that for a fixed radiative efficiency, {\it the quasar lifetime
is independent of redshift} while the Hubble time is shorter at high
redshift.  For a flat universe, the ratio between $t_{\rm Edd}$ and the
Hubble time $H^{-1}$ is, $\sim 10\% [(1+z)/10]^{3/2} \epsilon_{0.1}{\tilde
L}^{-1}$.  Therefore several tenths of all star clusters might show nuclear
activity at $z\sim 10$. Based on these numbers we infer that quasars would
dominate the UV background during reionization if the fraction of gas
converted to black holes is $\ga 10^{-3}\epsilon_{0.1}^{-1}$.

\begin{figure}
\includegraphics{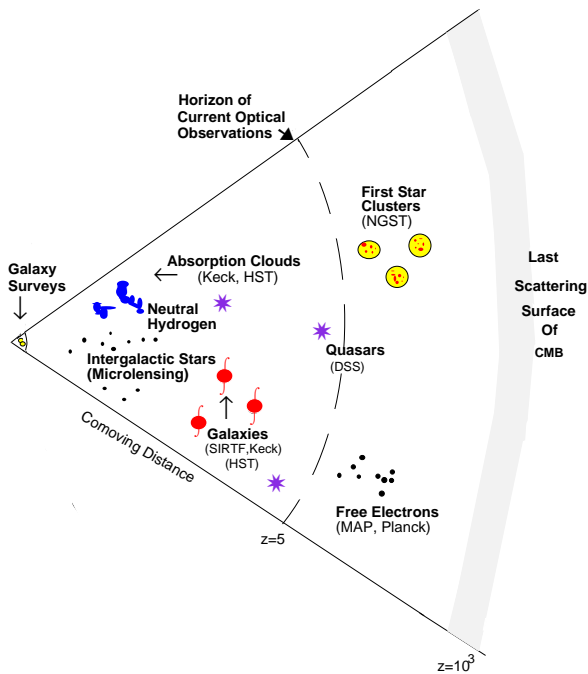}
\vspace{3.1in}
\caption{Summary: the zoo of different objects that populate the high-redshift 
universe. Listed below each entry are related instruments or observational
programs. The Sloan Digital Sky Survey (DSS; see
http://www.astro.princeton.edu/BBOOK), will catalog $\sim 10^5$ quasars, an
order of magnitude more than the number in current surveys.  
}
\label{fig-5}
\end{figure}

We have suggested that the formation efficiency of black holes in low-mass
galaxies at $z\sim 10$ is similar to that found in galaxies in the local
Universe.  It is however possible that the efficiency of black hole
formation is substantially reduced in the shallow potential wells of the
early population of sub-galactic halos (Haehnelt \& Rees 1993).
Observations with NGST will determine whether this is the case or
whether quasars indeed pre-dated massive galaxies (Loeb 1995).

\acknowledgements

I thank Zoltan Haiman, Kim McLeod, Jerry Ostriker, and Martin Rees for
useful discussions, and Alexandra Schmitz for assistance in the production
of the figures.  This work was supported in part by the NASA ATP grant
NAG5-3085 and the Harvard Milton fund.

\end{document}